\documentclass[a4paper, 11pt]{article}
\usepackage{amsmath}
\usepackage{graphicx}
\usepackage{latexsym}

\def\b{\begin{eqnarray}}
\def\e{\end{eqnarray}}
\def\n{\noindent}

\newtheorem{proposition}{Proposition}
\newtheorem{definition}{Definition}

\begin{document}

\begin{center}

{\LARGE\textbf{Conformal Properties and B{\"a}cklund Transform for
the Associated Camassa-Holm Equation \\}} \vspace {10mm}
\vspace{1mm} \noindent

{\large \bf Rossen Ivanov$^\ast$}\footnote{On leave from the
Institute for Nuclear Research and Nuclear Energy, Bulgarian Academy
of Sciences, Sofia, Bulgaria.} \vskip1cm \n \hskip-.3cm
\begin{tabular}{c}
\hskip-1cm $\phantom{R^R}${\it School of Mathematics, Trinity
College,}
\\ {\it Dublin 2, Ireland} \\ {\it Tel:  + 353 - 1 - 608 2898 }\\{\it  Fax:  + 353 - 1- 608 2282} \\
\\ {\it $^\ast$e-mail: ivanovr@maths.tcd.ie} \\
\\
\hskip-.8cm
\end{tabular}
\vskip1cm
\end{center}

\vskip1cm

\begin{abstract}
\n Integrable equations exhibit interesting conformal properties and
can be written in terms of the so-called conformal invariants. The
most basic and important example is the KdV equation and the
corresponding Schwarz-KdV equation.  Other examples, including the
Camassa-Holm equation and the associated Camassa-Holm equation are
investigated in this paper. It is shown that the B{\"a}cklund
transform is related to the conformal properties of these equations.
Some particular solutions of the Associated Camassa-Holm Equation
are discussed also.



\end{abstract}

\newpage

\section{Introduction}

\n Integrable equations exhibit many extraordinary properties like
infinitely many conservation laws, multi- Hamiltonian structures,
soliton solutions etc.  Many integrable equations in {1+1}
dimensions like KdV, MKdV, Harry-Dym, Boussinesq equations possess
interesting conformal properties as well \cite{D03, D87, W88, DS95,
Lou}. They can be written in terms of the so-called independent
conformal invariants of the function $\phi=\phi(x,t)$:

\b \label{eq1}
p_{1} & = & \frac{\phi_{t}}{\phi_{x}}\nonumber \\
p_{2} & = & \{\phi;x\}\equiv\frac{\phi_{xxx}}{\phi_{x}}-\frac{3}{2}\frac{\phi_{xx}^{2}}{\phi_{x}^{2}}\nonumber \\
\e

\n Here $\{\phi;x\}$ denotes the Schwarz derivative. A quantity is
called conformally invariant if it is invariant under the M{\"o}bius
transformation

\b \label{eq2} \phi & \rightarrow&
\frac{\alpha\phi+\beta}{\gamma\phi+\delta}, \qquad
\alpha\delta\neq\beta\gamma. \e

\n For example, the KdV equation \b \label{eq3}
u_{t}+au_{xxx}+3uu_{x}=0
 \e
\n ($a$ is a constant) can be written in a Schwarzian form, i.e. in
terms of the conformal invariants (\ref{eq1}) as $p_{1}+ap_{2}=0$ or

\b \label{eq3b} \frac{\phi_{t}}{\phi_{x}}+a\{\phi;x\}=0 \e

\n where

\b \label{eq3a} u=a\{\phi;x\}. \e

The KdV and Camassa-Holm (CH) \cite{CH93} equations arise also as
equations of the geodesic flow for the $L^{2}$ and $H^{1}$ metrics
correspondingly on the Bott-Virasoro group \cite{M98, CK03, CKKT04}.
The conformal properties of these equations and their link to the
Bott-Virasoro group originate from the Hamiltonian operator

\b \label{eq4} D_{3}\equiv a\partial^{3}+\partial u(x) +u(x)\partial
\e

\n where $\partial\equiv \partial/\partial x$. $D_{3}$ is also known
as the third Bol operator \cite{B49} and is the conformally
covariant version of the differential operator $\partial^{3}$, see
also \cite{G00}. $D_{3}$ defines a Poisson bracket \cite{D03}

\b \label{eq5} \{f,g\}=\frac{1}{2\pi
i}\int\Big((a\partial^{3}+\partial u(x) +u(x)\partial)\frac{\delta
f}{\delta u}\Big). \frac{\delta g}{\delta u} dx\e

\n and KdV, CH and many other equations \cite{FOR96} can be written
in the form

\b \label{eq6}  u_{t}=\{u,H\}\e

\n for some Hamiltonian $H$.  On the other hand, suppose for
simplicity that $u$ is $2\pi$ periodic in $x$, i.e.

\b \label{eq7}
u(x)=\sum^{\infty}_{-\infty}L_{n}e^{inx}-\frac{a}{2}\e

\n then the Fourier coefficients $L_{n}$ close a classical Virasoro
algebra with respect to the Poisson bracket (\ref{eq5}):

\b \label{eq8}
\{L_{n},L_{m}\}=(n-m)L_{n+m}+a(n^{3}-n)\delta_{n+m,0}\e

\n Therefore, it is natural to expect that equations which can be
written in the form (\ref{eq6}) exibit interesting conformal
properties, and this is the case indeed \cite{Lou, G00, W83, WTC83,
N89}. In what follows we will concentrate on the Camassa-Holm and
the associated Camassa-Holm equation.

\section{The Camassa-Holm equation in Schwarzian form}

The Camassa-Holm equation \cite{CH93, FF81}
\begin{equation}\label{eq9}
 u_{t}-u_{xxt}+a u_{xxx} +3uu_{x}-2u_{x}u_{xx}-uu_{xxx}=0,
\end{equation}
describes the unidirectional propagation of shallow water waves over
a flat bottom \cite{CH93, J02}. CH is a completely integrable
equation \cite{BBS98, CM99, C01, L02}, describing permanent and
breaking waves \cite{CE98, C00}. Its solitary waves are stable
solitons \cite{BBS99, CS00, CS02, J03}. CH arises also as an
equation of the geodesic flow for the $H^{1}$ metrics on the
Bott-Virasoro group \cite{M98, CK03, CKKT04}. The equation
(\ref{eq9}) admits a Lax pair \cite{CH93}

\b \label{eq10}
V_{xx}&=&\Big(\frac{1}{4}-\lambda(m+\frac{a}{2})\Big)V
 \\\label{eq11}
V_{t}&=&-\Big(\frac{1}{2\lambda}+u-a\Big)V_{x}+\frac{u_{x}}{2}V \e

 \n where
\b\label{eq12} m = u-u_{xx}. \e

\n In order to find the Schwarz form for the CH equation we proceed
as follows. Let $V_{1}$ and $V_{2}$ be two linearly independent
solutions of the system (\ref{eq10}), (\ref{eq11}) and let us define

\b \label{eq13} \phi=\frac{V_{2}}{V_{1}} \e

\n Then, from (\ref{eq11}) it follows that

\b \label{eq14}
\frac{\phi_{t}}{\phi_{x}}=-u+\Big(a-\frac{1}{2\lambda}\Big) \e

\n According to the Theorem 10.1.1 from \cite{H76} due to
(\ref{eq10}) we also have \b \label{eq15} \{\phi; x\}=2\lambda m
+a\lambda -\frac{1}{2} \e

\n From (\ref{eq12}), (\ref{eq14}) and (\ref{eq15}) we obtain the
Schwarz-Camassa-Holm (S-CH) equation:

\b \label{eq16}
(1-\partial^{2})\frac{\phi_{t}}{\phi_{x}}+\frac{1}{2\lambda}\{\phi;x\}=\frac{3}{2}a-\frac{3}{4\lambda}
\e

\n Since $\lambda$ is an arbitrary constant, one can take
$\lambda=1/2a $ (if $a\neq0$) and then the S-CH equation
(\ref{eq16}) acquires the form $(1-\partial^{2})p_{1}+ap_{2}=0$ or

\b \label{eq17}
(1-\partial^{2})\frac{\phi_{t}}{\phi_{x}}+a\{\phi;x\}=0 \e

Applying the hodograph transform

\b \label{eq17a} x\rightarrow \phi, \qquad  t\rightarrow t, \qquad
\phi \rightarrow x \e

\n to the S-CH (\ref{eq17}) and using the transformation properties
of the Schwarzian derivative \cite{H76}

\b \label{eq17b} \{\phi;x\}=-\phi_{x}^{2}\{x;\phi\} \e

\n we obtain the following integrable deformation of the Harry Dym
equation for the variable $v=1/x_{\phi}$:

\b \label{eq17c}
v_{t}+v^{2}[v(v\partial_{\phi}^{-1}(v^{-1})_{t})_{\phi}]_{\phi}=av^{3}v_{\phi\phi\phi}
\nonumber \e

The equations (\ref{eq9}) and (\ref{eq17}) with
$u=-\frac{\phi_{t}}{\phi_{x}}$ are not equivalent -- as a matter of
fact (\ref{eq17}) implies (\ref{eq9}), cf. \cite{W88}.  It is often
convenient to think that the Lax operator belongs to some Lie
algebra, and the corresponding Jost solution- to the corresponding
group. Thus the relation between $u$ and $\phi$ (see (\ref{eq13}))
resembles the relation between the Lie group and the corresponding
Lie algebra, as pointed out in \cite{W88}. More precisely, the
following proposition holds:

\begin{proposition}\label{p1}
Let $\phi$ be a solution of (\ref{eq16}). Then
$V_{1}=\phi^{-1/2}_{x}$ and $V_{2}=\phi \phi^{-1/2}_{x}$ are
solutions of (\ref{eq10}), (\ref{eq11}) with $
u=-\frac{\phi_{t}}{\phi_{x}}-(\frac{1}{2\lambda}-a)$
\end{proposition}
{\it Proof }:  It follows easily by a direct computation.

\bigskip

Note that the Proposition \ref{p1} is consistent with (\ref{eq13}).
From Proposition \ref{p1} it follows
\begin{proposition}
The general solution of (\ref{eq10}), (\ref{eq11}) is \b
\label{eq18} V=\frac{A\phi+B}{\sqrt{\phi_{x}}} \e \n where $A$ and
$B$ are two arbitrary constants, not simultaneously zero.
\end{proposition}

Note that the expression (\ref{eq18}) is covariant with respect to
the M{\"o}bius transformation (\ref{eq2}), i.e. under (\ref{eq2}),
the expression (\ref{eq18}) transforms into an expression of the
same form but with constants

\b \label{eq19} A\rightarrow A'=\frac{\alpha A + \gamma
B}{\sqrt{\alpha\delta-\beta\gamma}},\qquad B\rightarrow
B'=\frac{\beta A + \delta B}{\sqrt{\alpha\delta-\beta\gamma}}. \e

\section{The associated Camassa-Holm equations and the B{\"a}cklund Transform}

An inverse scattering method, which can be applied directly to the
spectral problem (\ref{eq10}) is not developed completely yet.
However, the solutions of the CH (in parametric form) can be
obtained by an implicit change of variables, which maps (\ref{eq10})
to the well known  spectral problem of the KdV hierarchy. This
change of variables leads to the so-called Associated Camassa-Holm
equation (ACH) \cite{S98}

\b \label{eq20} p_{t'}=p^{2}f _{x'}, \qquad f=\frac{p}{4} (\log
p)_{x't'}-\frac{p^{2}}{2}. \e

\n Indeed, the ACH is related to the CH equation (\ref{eq9}) through
the following change of variables:

\b \label{eq21} p=\sqrt{m+\frac{a}{2}}, \qquad
f=-\frac{u}{2}-\frac{a}{4} \e

\b \label{eq22} dx'=\frac{1}{2}p dx+ (-\frac{u}{2}+\frac{a}{2})p dt,
\qquad dt'=dt. \e

\n In what follows we will omit the dash from the new variables $x'$
and $t'$ and we will write simply $x$ and $t$ instead.

The ACH equation (\ref{eq20}) is related to the KdV hierarchy
\cite{FS99, H99} since it may be written in the form

\b \label{eq23} U_{t}=-2p_{x}, \qquad
U=-\frac{1}{2}\Big(\frac{pp_{xx}-\frac{1}{2}p_{x}^{2}+2}{p^{2}}\Big)
\e

\n and the Lax pair for (\ref{eq23}) is

\b \label{eq24} V_{xx}+\Big(U-\frac{1}{\lambda}\Big)V=0,
 \\\label{eq25}
V_{t}=\lambda\Big(pV_{x}-\frac{1}{2}p_{x}V\Big). \e

\n  The second part of (\ref{eq23}) is the Ermakov-Pinney equation
for $p$ (for a known $U$) \cite{H99, C01, J03, Sh05}. In order to
obtain the Schwarzian form of the ACH equation, we introduce again
$\phi=V_{2}/V_{1}$ as in (\ref{eq13}), where now $V_{1}$ and $V_{2}$
are two linearly independent solutions of the system (\ref{eq24}),
(\ref{eq25}). Then, by analogous arguments, from (\ref{eq24}) it
follows

\b \label{eq26} U=\frac{1}{2} \{\phi;x\}+\frac{1}{\lambda},\e

\n and from (\ref{eq25})

\b \label{eq27} p=\frac{\phi_{t}}{\lambda \phi_{x}}.\e

\n Using the relation between $p$ and $U$ given in the first part of
(\ref{eq23}), (\ref{eq26}) and (\ref{eq27}) we obtain the
Schwarz-ACH (S-ACH) equation $p_{2,t}+(4/\lambda)p_{1,x}=0$, or

\b \label{eq28}
\{\phi;x\}_{t}+\frac{4}{\lambda}\Big(\frac{\phi_{t}}{
\phi_{x}}\Big)_{x}=0. \e

Now one can easily check by a direct computation the validity of the
following

\begin{proposition}
Let $\phi$ be a solution of S-ACH (\ref{eq28}). Then
$V_{1}=\phi^{-1/2}_{x}$ and $V_{2}=\phi \phi^{-1/2}_{x}$ are
solutions of (\ref{eq24}), (\ref{eq25}) with \b \label{eq28a}
p=\frac{\phi_{t}}{\lambda \phi_{x}}.\e
\end{proposition}

As a corollary we find that the general solution of (\ref{eq24}),
(\ref{eq25}), expressed through the solution of (\ref{eq28}) is

\b \label{eq29} V=\frac{A\phi+B}{\sqrt{\phi_{x}}} \e

\n where $A$ and $B$ are two arbitrary constants, not simultaneously
zero. As we know the expression (\ref{eq29}) is covariant under the
M{\"o}bius transformations (\ref{eq2}).

\begin{definition} \nonumber
A transformation, which connects the solution $u$ of one equation to
the solution $\widetilde{u}$ of the same (or another) equation is
called a B{\"a}cklund Transform (BT).
\end{definition}

\n Such type of transformations were first discovered in relation to
the sine-Gordon equation by A. B{\"a}cklund around 1883. Usually BT
depends on an arbitrary parameter which is called B{\"a}cklund
parameter.

The Lax pair (\ref{eq24}), (\ref{eq25}) has the following important
property \cite{H99,HKR99}, namely the BT for the ACH :

\begin{proposition}
If $V$ is a solution of the Lax pair (\ref{eq24}), (\ref{eq25}) with
potentials $U$ and $p$, then $V^{-1}$ is a solution of (\ref{eq24}),
(\ref{eq25}) with potentials

\b \label{eq30} \widetilde{U}&=&U+2(\log V)_{xx},\\\label{eq31}
\widetilde{p}&=&p-(\log V)_{xt}.\e
\end{proposition}

From (\ref{eq29}) and (\ref{eq31}) it now follows that the BT for
the solution (\ref{eq28a}) of the ACH equation can be expressed
through the solution of the S-ACH equation as follows:

\b \label{eq32} \widetilde{p}=\frac{\phi_{t}}{\lambda
\phi_{x}}-\Big(\log \frac{A\phi+B}{\sqrt{\phi_{x}}}\Big)_{xt}.\e

Since the constants $A$, $B$ are not simultaneously zero, one can
always factor out one of the two constants, i.e.  (\ref{eq32})
obviously depends only on the ratio of these two constants (which is
the B{\"a}cklund parameter in this case).

We conclude this section with the following remark for the KdV
equation. It is not difficult to see through similar considerations,
that the B{\"a}cklund Transform for the solution (\ref{eq3a}) of the
KdV equation (\ref{eq3}) is \b \label{eq33a}
\widetilde{u}=a\{\phi;x\}+4a \Big(\log
\frac{A\phi+B}{\sqrt{\phi_{x}}}\Big)_{xx},\e

\n where $\phi$ is the solution of the S-KdV equation (\ref{eq3b}).
Then the choice $A=0$ in (\ref{eq33a}) corresponds to the 'exotic'
B{\"a}cklund Transform found by Galas in \cite{G92} and mentioned in
\cite{S96}.

\section{On the solutions of the ACH and S-ACH equations}

The solutions of the ACH and S-ACH equations can be constructed,
following the scheme for the construction of the solutions of the
KdV hierarchy \cite{H99, MS91, M92a, M92b, M94, R96}.

Let $v_{1}, v_{2},\ldots v_{n}$ be $n$ different solutions of the
Lax pair (\ref{eq24}), (\ref{eq25}) with potentials $U$ and $p$,
taken at $\lambda= \lambda_{1}, \lambda_{2},\ldots \lambda_{n}$
respectively and let $V_{1}$ and $V_{2}$ be two linearly independent
solutions for some value $\lambda$.  Consider the Wronskian
determinants

\b  W&=&W(v_{1},v_{1;1},\ldots, v_{1;m_{1}}, v_{2}, v_{2;1},\ldots,
v_{2;m_{2}},\ldots,  v_{n},v_{n;1},\ldots, v_{n;m_{n}}) \nonumber
 \\
W_{k}&= &W(v_{1},v_{1;1},\ldots, v_{1;m_{1}}, \ldots,
v_{n},v_{n;1},\ldots, v_{n;m_{n}},V_{k}), \qquad k=1,2,\nonumber
\\\label{eq33} \e

\n where $v_{i;l}\equiv \partial_{\lambda}^{l}v_{i}$,  $m_{i}$ are
arbitrary nonnegative integers and the Wronskian determinant of $N$
functions $\varphi_{1}, \varphi_{2},\ldots \varphi_{N}$ is defined
by

\b  \label{eq34} W(\varphi_{1}, \varphi_{2},\ldots \varphi_{N})=\det
A, \qquad A_{ij}=\frac{d^{i-1}\varphi_{j}}{dx^{i-1}}, \qquad
i,j=1,2,\ldots,N. \e

\n The following generalization of the Crum theorem is valid for the
KdV hierarchy and in particular for the ACH equation \cite{M92a}:

\begin{proposition}\label{p4}
$\widetilde{V_{k}}=W_{k}/W$, $k=1,2$ are two linearly independent
solutions of the Lax pair (\ref{eq24}), (\ref{eq25}) with potentials
\b \label{eq35} \widetilde{U}&=&U+2(\log W)_{xx},\\ \label{eq36}
\widetilde{p}&=&p-(\log W)_{xt}.\e
\end{proposition}

\n From Proposition \ref{p4} it follows immediately that \b
\label{eq37}
\widetilde{\phi}=\frac{\widetilde{V_{2}}}{\widetilde{V_{1}}}=\frac{W_{2}}{W_{1}},
\qquad \phi=\frac{V_{2}}{V_{1}}\e are the solutions of the S-ACH
(\ref{eq28}), which correspond to the potentials \b
\widetilde{p}=\frac{\widetilde{\phi_{t}}}{\lambda
\widetilde{\phi_{x}}}, \qquad  p=\frac{\phi_{t}}{\lambda \phi_{x}}.
\e

In addition to the solutions mentioned in \cite{H99} one may
construct also the so-called negaton and positon solutions (which
are singular) and mixed soliton-positon-negaton solutions. For a
detailed discussion on the positon and negaton solutions of KdV we
refer to \cite{R96}.

Let us return to (\ref{eq24}) with $p=h=\mathrm{const}$, i.e.
$U=-1/h^2$. Then it has the form

\b\label{eq280} V_{xx}+E(h,\lambda)V=0, \qquad
E(h,\lambda)=-\Big(\frac{1}{h^2}+\frac{1}{\lambda}\Big).\e

\n The type of the solutions of (\ref{eq280}) clearly depends on the
sign of the constant $E(h,\lambda)$. If $E<0$ the independent
solutions of (\ref{eq280}) and (\ref{eq25}) are

\b\label{eq281} v_{1} (\lambda)& =&
\cosh\Big(\sqrt{\frac{1}{\lambda}+\frac{1}{h^{2}}}(x-h\lambda
t+x_{1}(\lambda))\Big),
\\ \label{eq282}v_{2} (\lambda) &=&\sinh \Big(\sqrt{\frac{1}{\lambda}+\frac{1}{h^{2}}}(x-h\lambda t+x_{2}(\lambda))\Big),\e

\n where $x_{1,2}(\lambda)$ are arbitrary functions. The simplest
choice for $W$ in (\ref{eq35}), (\ref{eq36}) is  $W=v_1$ or $W=v_2$
(Wronskian of order one), which corresponds to the {\it one-soliton
solution}.  The case $W=W(v_1(\lambda_1),v_2(\lambda_2))$
corresponds to the {\it two-soliton solution}, i.e. this solution
describes the interaction of two solitons. However, according to
(\ref{eq33}) it is possible to merge two or more identical solitons,
taking

\b  \label{eq283} W&=&W(v_{1},v_{1;1},\ldots, v_{1;m_{1}}), \e

\n or \b \label{eq284} W&= &W(v_{2},v_{2;1},\ldots, v_{2;m_{2}}). \e

In physical terms $E(h,\lambda)$ is the energy of the soliton. Since
the energy of the solitons is negative, (\ref{eq283}) is called {\it
negaton} of type $C$ and order $m_1$, and (\ref{eq284}) is called
{\it negaton} of type $S$ and order $m_2$, cf. \cite{R96}. The
analogous considerations resulting from the following solutions of
(\ref{eq280}),

\b\label{eq285} v_{1} (\lambda) &=&
\cos\Big(\sqrt{\frac{1}{\lambda}-\frac{1}{h^{2}}}(x-h\lambda
t+x_{1}(\lambda))\Big),
\\ \label{eq286}v_{2} (\lambda) &=&\sin \Big(\sqrt{\frac{1}{\lambda}-\frac{1}{h^{2}}}(x-h\lambda t+x_{2}(\lambda))\Big)\e

\n ($0<\lambda\leq h^{2}$), whose energy is positive, $E>0$ leads
naturally to the definition of {\it positon} of type $C$ and order
$m_1$ via (\ref{eq283}) and (\ref{eq36}), and {\it positon} of type
$S$ and order $m_2$, (\ref{eq284}) and (\ref{eq36}); cf. \cite{R96}.

Due to (\ref{eq33}) it is possible to construct solutions, which
describe the interaction between solitons, positons and negatons.
I.e. the interaction between $n$ positons (of orders
$m_1,m_2,\ldots,m_n$) and $N$ solitons is given by the { \it
$n$-positon -- $N$ soliton solution } $_{n}p_{N}$ of the ACH
(\ref{eq20}) which can be constructed from (\ref{eq36}) taking the
following solutions of the Lax pair (\ref{eq24}), (\ref{eq25}): \b
\label{eq38}
v_{i}=\cos\Big(\sqrt{\frac{1}{\lambda_{i}}-\frac{1}{h^{2}}}
(x-h\lambda_{i}t+y_{i}(\lambda_{i}))\Big),\qquad i=1,\ldots n \\
\label{eq39}  \eta_{j}  =
\exp\Big(\sqrt{\frac{1}{\lambda_{n+j}}+\frac{1}{h^{2}}}(x-h\lambda_{n+j}t+x_{j})\Big)+
\qquad \qquad \qquad \qquad
\\c_{j}\exp\Big(-\sqrt{\frac{1}{\lambda_{n+j}}+\frac{1}{h^{2}}}(x-h\lambda_{n+j}t+x_{j})\Big),\qquad j=1,\ldots N,\e

\n where $y_{i}(\lambda)$ are arbitrary functions of $\lambda$,
$c_{j}$, $x_{j}$ are arbitrary constants and $0<\lambda_{i}\leq
h^{2}$ for $i=1,\ldots,n$. Then (cf. \cite{M94})

\b \label{eq40} _{n}p_{N}(x,t)&=&\nonumber\\
h\!\!\!\!&-&\!\!\!\!\Big(\log W(v_{1},v_{1;1},\ldots, v_{1;m_{1}},
\ldots, v_{n},v_{n;1},\ldots,
v_{n;m_{n}},\eta_{1},\ldots,\eta_{N})\Big)_{xt}.\nonumber\\ \e

\begin{figure}
\centering
\includegraphics[height=8cm]{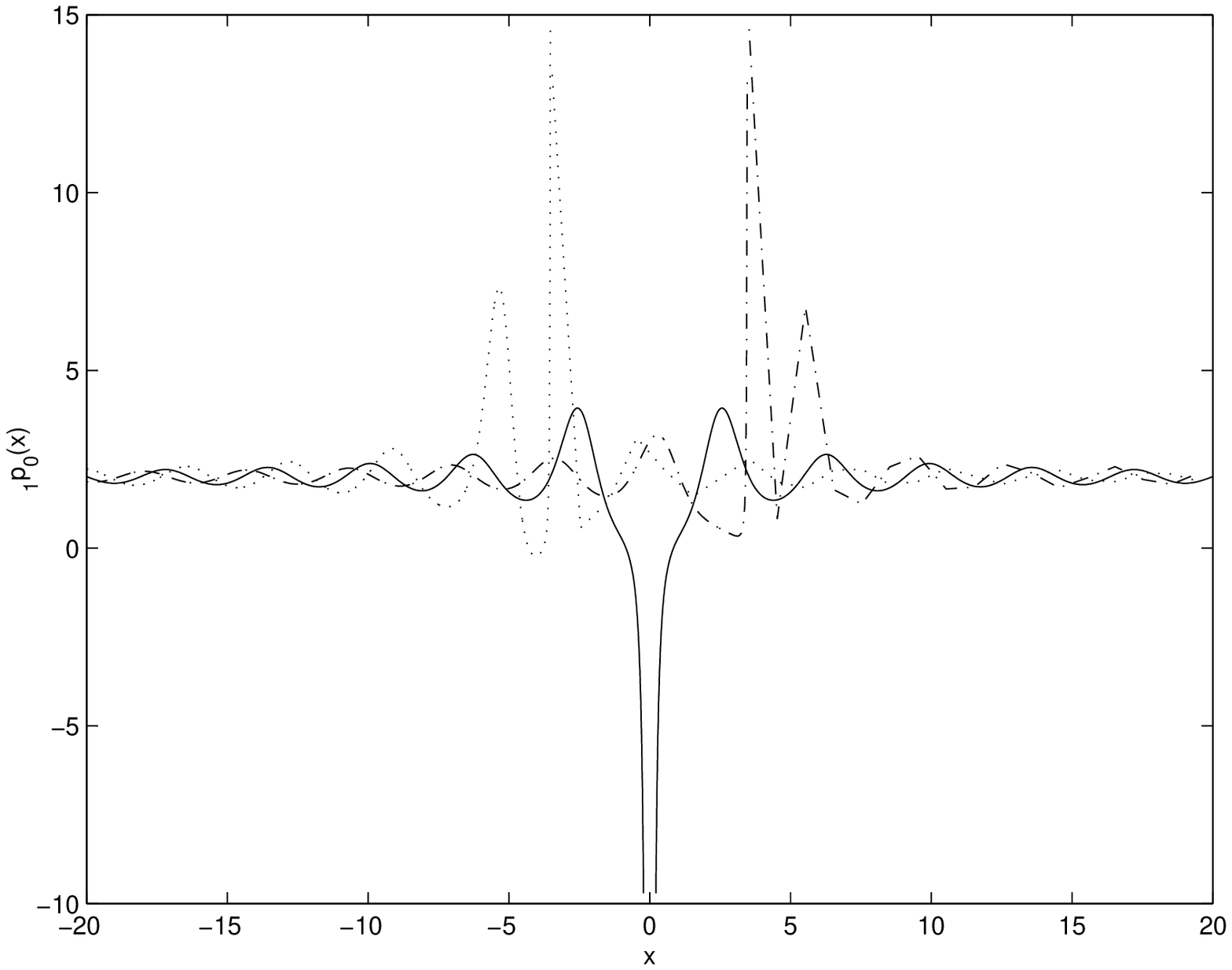}
%
%
\caption{ One-positon solution of the ACH equation, $a=0$, $h=2$,
$\lambda_{1}=1$: $t=0$ -- solid line; $t=-4$ -- dashed line; $t=4$
-- dotted line. }
\label{1positon}       
\end{figure}

As an example, we provide the one-positon solution of order one for
the ACH, see also Fig. \ref{1positon}: \b \label{eq41}
_{1}p_{0}(x,t)=h-\frac{8h(h^{2}-\lambda_{1})\Big(\lambda_{1}+\lambda_{1}\cos
z(x,t) + r(x,t)\sin z(x,t)\Big)}{\Big(2r(x,t)+h^{2}\sin
z(x,t)\Big)^{2}} \e

\n where $\omega=\sqrt{\frac{1}{\lambda_{1}}-\frac{1}{h^{2}}}$,
$y_{1}=a\lambda$, $a$ is an arbitrary constant;

\b \label{eq42} z(x,t)&=&2\omega (x-h\lambda_{1}t+a \lambda_{1}), \nonumber\\
 r(x,t)&=&\omega \Big(h^{2}x+\lambda_{1}(h^{2}-2\lambda_{1})(ht-a)\Big).\nonumber\e

\section{Acknowledgements}

The author is grateful to Prof. A. Constantin for helpful
discussions and to the referees for valuable suggestions. The author
also acknowledges funding from the Science Foundation Ireland, Grant
04/BR6/M0042.

\newpage

\end{document}